\documentclass[article,nojss]{jss}

\usepackage{thumbpdf,lmodern}

\usepackage{framed}
\usepackage{comment}

\usepackage[utf8]{inputenc}

\usepackage{amsmath}
\usepackage{amsfonts}
\usepackage{mathrsfs}
\usepackage[dvipsnames]{xcolor}
\newcommand\Tstrut{\rule{0pt}{2.5ex}}

\author{Guilherme Bodin\\PUC-Rio\\Rio de Janeiro, Brazil
   \And Raphael Saavedra\\Invenia Labs\\Cambridge, UK
   \AND Cristiano Fernandes\\PUC-Rio\\Rio de Janeiro, Brazil
   \And Alexandre Street\\PUC-Rio\\Rio de Janeiro, Brazil}
\Plainauthor{Guilherme Bodin, Raphael Saavedra, Cristiano Fernandes, Alexandre Street}

\title{\pkg{ScoreDrivenModels.jl}: a \proglang{Julia} Package for \\ Generalized Autoregressive Score Models}
\Plaintitle{ScoreDrivenModels.jl: Generalized Autoregressive Score Models in Julia}
\Shorttitle{\pkg{ScoreDrivenModels.jl}: Generalized Autoregressive Score Models in \proglang{Julia} }

\Abstract{
 Score-driven models, also known as generalized autoregressive score (GAS) models, represent a class of observation-driven time series models. They possess powerful properties, such as the ability to model different conditional distributions and to consider time-varying parameters within a flexible framework. In this paper, we present \pkg{ScoreDrivenModels.jl}, an open-source \proglang{Julia} package for modeling, forecasting, and simulating time series using the framework of score-driven models. The package is flexible with respect to model definition, allowing the user to specify the lag structure and which parameters are time-varying or constant. It is also possible to consider several distributions, including Beta, Exponential, Gamma, Lognormal, Normal, Poisson, Student's t, and Weibull. The provided interface is flexible, allowing interested users to implement any desired distribution and parametrization.
}

\Plainkeywords{score-driven models, generalized autoregressive score models, time series models, time-varying parameters, non-Gaussian models}

\Address{
  Guilherme Bodin (corresponding author)\\
  Department of Electrical Engineering\\
  Pontifícia Universidade Católica do Rio de Janeiro\\
  R. Marquês de São Vicente, 225\\
  Gávea, Rio de Janeiro - RJ, Brazil\\
  E-mail: \email{guilherme.b.moraes@gmail.com}\\
  \\
  Raphael Saavedra\\
  Invenia Labs\\
  95 Regent St\\
  Cambridge CB2 1AW, United Kingdom\\
  E-mail: \email{raphael.saavedra@invenialabs.co.uk}\\
  \\
  Cristiano Fernandes\\
  Department of Electrical Engineering\\
  Pontifícia Universidade Católica do Rio de Janeiro\\
  R. Marquês de São Vicente, 225\\
  Gávea, Rio de Janeiro - RJ, Brazil\\
  E-mail: \email{cris@ele.puc-rio.br}\\
  \\
  Alexandre Street\\
  Department of Electrical Engineering\\
  Pontifícia Universidade Católica do Rio de Janeiro\\
  R. Marquês de São Vicente, 225\\
  Gávea, Rio de Janeiro - RJ, Brazil\\
  E-mail: \email{street@ele.puc-rio.br}\\
}

\begin{document}

\section[Introduction]{Introduction} \label{sec:intro}

Time series models with time-varying parameters have become increasingly popular over the years due to their advantages in capturing dynamics of series of interest. According to \cite{cox1981parameterdriven}, the mechanism driving parameter dynamics in this general class of models can be of two types: parameter-driven, as in state space models \citep{durbin2012time, koopman2000stamp, saavedra2019statespacemodels}, or observation-driven. In this work, we will focus on a recently proposed class of observation-driven models wherein the score of the predictive density is used as the driver for parameter updating \citep{creal2013generalized, harvey2013dynamic}. These models have been referred to as generalized autoregressive score (GAS) models, dynamic conditional score models, or simply score-driven models. Additionally, it has been demonstrated that well-established observation driven models, such as the GARCH \citep{bollerslev1986generalized} and conditional duration models \citep{engle1998duration}, are particular cases of the score-driven framework.

Two of the main advantages of the GAS framework are its ability to consider different non-Gaussian distributions and its flexibility with respect to the updating mechanism, which is determined by the chosen distribution determines the model updating mechanism. These properties have led GAS models to be applied in numerous fields, such as finance \citep{harvey2016testing, ayala2018score}, actuaries \citep{Neves2017, demelo2018forecasting}, risk analysis \citep{patton2019dynamic, nani2019value}, and renewable generation \citep{saavedra2017study, Hoeltgebaum2018}. We also refer the interested reader to a large online repository of works on GAS models at \url{http://www.gasmodel.com}. This wide range of applications has motivated the development of software packages for this class of models. For instance, there are open-source packages in \proglang{Python} \citep{pyfluxTaylor}, \proglang{R} \citep{Ardia2019}, and, recently, the data consultancy company Nlitn have made publicly available the Time Series Lab (\url{https://timeserieslab.com}), a free software developed by some of the authors of the theory developed in \citep{creal2013generalized, harvey2013dynamic}.

In this paper, we present a novel open-source GAS package fully implemented in \proglang{Julia} \citep{bezanson2017julia} named \pkg{ScoreDrivenModels.jl} \citep{bodin2020gas}. One of \proglang{Julia}'s main advantages is to avoid the so-called two-language problem, i.e., the dependence on subroutines implemented in lower-level languages such as \proglang{C}/\proglang{C++} or \proglang{Fortran}. \proglang{Julia} achieves this by providing a high-level programming syntax that allows for rapid prototyping and development without sacrificing computational performance. Thus, by providing an open-source package completely written in \proglang{Julia}, we facilitate development and contributions by users while also maintaining a high level of code transparency. The package allows users to specify a wide variety of GAS models by choosing the conditional distribution, the autoregressive structure, and which parameters are time-varying. Finally, initialization procedures are implemented to turn the estimation process more robust for the case of seasonal time series.


The remainder of this paper is organised as follows. Section \ref{sec:sdm} provides a brief overview of the GAS framework. In Section \ref{sec:package}, the \pkg{ScoreDrivenModels.jl} package is presented, including the model specification, estimation, forecasting, and simulation. Section \ref{sec:applications} presents examples of applications to illustrate the use of the package. Conclusions are drawn in Section \ref{sec:conclusion}. Finally, the Appendix provides the derivation of the score for each implemented distribution.



\section{Score-Driven Models} \label{sec:sdm}

\subsection{The GAS Framework}
\label{sec:framework}
Let $y_t \in \mathbb{Y} \subseteq \mathbb{R}$ denote the dependent variable of interest, $f_t \in \mathcal{P} \subset \mathbb{R}^k$ be a vector of time-varying parameters, and $Y^{t} = \{y_1, \dots, y_t\}$ and $F^{t} = \{f_0, f_1, \dots, f_t\}$ denote the sets of available information until time $t$. We assume that $y_t$ is generated by the probability density function conditioned on the available information (past data and time-varying parameters) and on the hyperparameter vector $\theta$, which contains the constant parameters. It follows that the predictive distribution of $y_{t}$ has a closed form, represented as:
\begin{equation}
    p(y_t| F^{t}, Y^{t-1}; \theta)
\end{equation}
In score-driven models, the updating mechanism for the time-varying parameters $f_t$ is given by the following equation, referred to as a GAS($p$, $q$) mechanism:
\begin{equation}
    f_{t+1} = \omega + \sum_{i=1}^p A_{i}s_{t-i+1} + \sum_{j=1}^q B_{j}f_{t-j+1},
    \label{recursion}
\end{equation}

where $\omega$ is a vector of constants, coefficient matrices $A_i$ and $B_j$ have appropriate dimensions for $i=1, \dots, p$ and $j=1, \dots, q$, and $s_t=s_t(y_t,f_t, F_{t-1}, Y_{t-1}; \theta)$ is an appropriate function of past data. The unknown coefficients in Eq. \eqref{recursion} are functions of the vector of hyperparameters $\theta$; that is, $\omega=\omega(\theta)$, $A_i=A_i(\theta)$, and $B_j=B_j(\theta)$. At instant $t$, the update of the time‐varying $f_t$ for the next period $t+1$ is conducted through Eq. \eqref{recursion}, with
\begin{equation}
s_{t} = \mathcal{I}_{t|t-1}^{-d} \cdot \nabla_{t}, \quad \quad
\nabla_{t} = \frac{\partial \ln p(y_{t} | f_t, F_{t-1}, Y_{t-1}; \theta)}{\partial f_{t}} ,
\end{equation}

where $\nabla_t$ is the called the score and $\mathcal{I}_{t|t-1}^{-d}$ is the scaled Fisher information of the probability density $p(y_{t} | f_{t}, \mathcal{F}_{t-1}; \theta)$. The scaling coefficient $d$ commonly takes values in $\{0, \frac{1}{2} , 1\}$. It is worth mentioning that in the case where $d = \frac{1}{2}$, it follows that $\mathcal{I}_{t|t-1}^{-\frac{1}{2}}$ results from the Cholesky decomposition of $\mathcal{I}_{t|t-1}^{-1}$.

\begin{figure}[h!]
\centering
\includegraphics[width=0.7\linewidth]{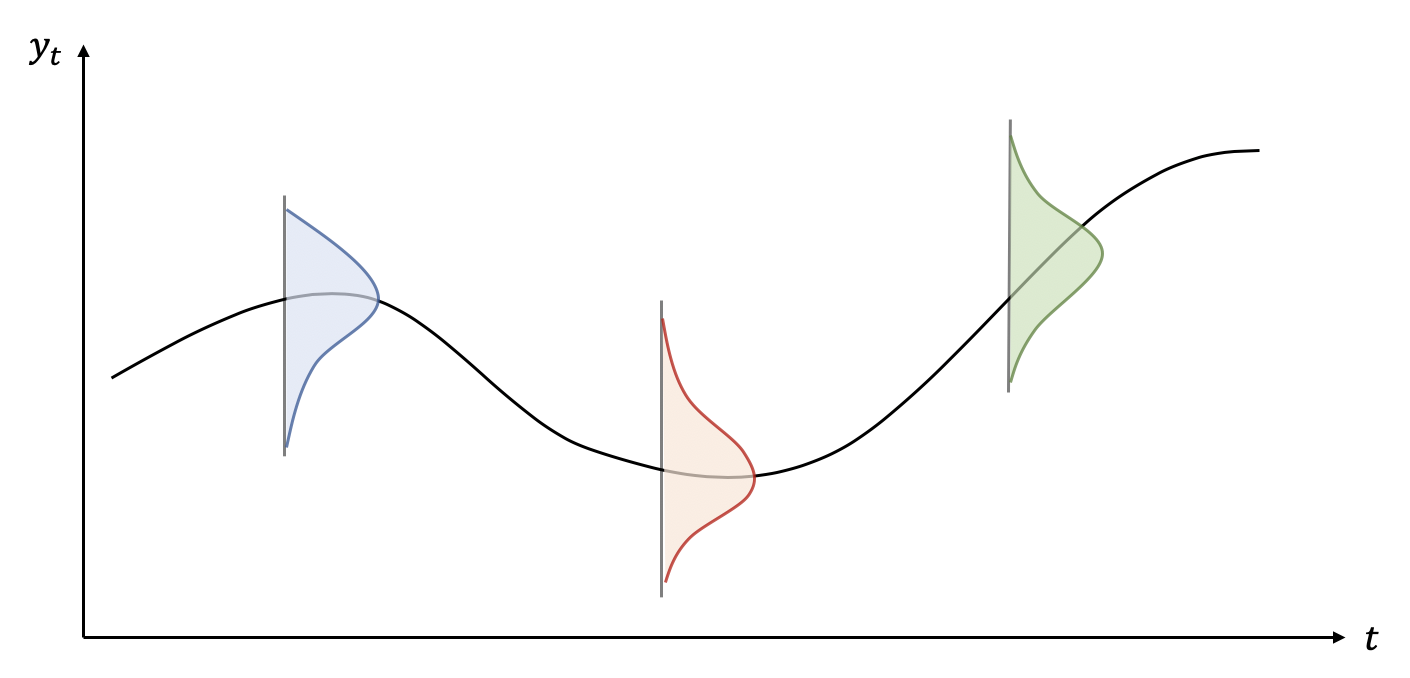}
\caption{\label{fig:Distributions} The GAS framework allows the conditional distribution to continuously change based on the data.}
\end{figure}

As a consequence of the time-varying mechanism for the distribution parameters presented in Eq. \eqref{recursion}, the conditional distribution of a GAS model is capable of continuously changing based on the considered data. For instance, if the time series contains occasional volatility spikes, the model can capture this behavior through the time-varying nature of the parameters. This property is illustrated in Fig. \ref{fig:Distributions}.

\subsection{Parametrization}
\label{sec:parametrization}

In the GAS updating mechanism \eqref{recursion}, the parameter $f_t \in \mathcal{P} \subset \mathbb{R}^k$ is sometimes bounded -- for example, in a Normal GAS model where the variance is time-varying, we would have $f_{t} = \sigma^2_t$ which can only assume positive values by definition. However, in some cases the recursion can lead to updates of $f_t \notin \mathcal{P}$. A solution is to reparametrize the equations in order to guarantee $f_t \in \mathcal{P}$ for every update. To that end, we follow the procedure described in \cite{creal2013generalized} and present an example to illustrate it.

Let $f_t$ be the vector of time-varying parameters of a Normal distribution. From the properties of the distribution, it follows that $\mu_t \in \mathbb{R}$ and $\sigma^{2}_{t} \in \mathbb{R}^+$. Let us define a new time-varying parameter $\tilde f_t \in \mathbb{R}^k$ and a map $h:\mathcal{P}\to\mathbb{R}^k$, which we denote the \code{link} function in the package. In the case of the Normal distribution, a useful approach is to have an \code{IdentityLink} for $\mu_t$ and a \code{LogLink} for $\sigma^2_t$ as follows:
\begin{equation}
    f_{t} = \begin{bmatrix} 
    \mu_t\\
    \sigma^2_t
    \end{bmatrix}, \tilde f_{t} = \begin{bmatrix} 
    \mu_t\\
    \ln \sigma^2_t
    \end{bmatrix}
\end{equation}

Note that the use of this parametrization will affect the recursion in Eq. \eqref{recursion} as well as the final expressions of $s_t$. Thus, let us derive the new recursion for the \eqref{recursion}, but this time with the guarantee that every update of $f_t$ respects $f_t \in \mathcal{P}$. To that end, we also define the inverse map $h^{-1}(\cdot)$ denoted in the software as the \code{unlink} function: $f_{t} = h^{-1}(\tilde{f}_{t})$. Then, we can define the GAS updating recursion utilizing the parametrization:
\begin{equation} \label{eq:param_recursion}
    \tilde{f}_{t+1} = \omega + \sum_{i=1}^p A_{i} \tilde{s}_{t-i+1} + \sum_{j=1}^q B_{j} \tilde{f}_{t-j+1}
\end{equation}

The unknown coefficients in \eqref{eq:param_recursion} remain as functions of $\theta$; that is, $\omega=\omega(\theta)$, $A_i=A_i(\theta)$, and $B_j=B_j(\theta)$. However, this time, we update the linked version of the time-varying parameters $\tilde{f}_t$ using \eqref{eq:param_recursion}. It is important to note that the expressions of $\tilde s_t$ are different for each scaling $d \in \{0, \frac{1}{2}, 1\}$. To compute these, we must use the derivative $\dot{h}$ of the map $h$, which is simply its Jacobian. In our example, $h$ is defined as
\begin{equation}
    h\left(\begin{bmatrix} 
    \mu_t\\
    \sigma^2_t
    \end{bmatrix}\right) = \begin{bmatrix} 
    h_1(\mu_t)\\
    h_2(\sigma^2_t)
    \end{bmatrix} = \begin{bmatrix} 
    \tilde \mu_t\\
    \tilde \sigma^2_t
    \end{bmatrix}= \begin{bmatrix} 
    \mu_t\\
    \ln \sigma^2_t
    \end{bmatrix}
\end{equation}
and its Jacobian is defined as 
\begin{equation}
    \dot h\left(\begin{bmatrix} 
    \mu_t\\
    \sigma^2_t
    \end{bmatrix}\right) = \begin{bmatrix} 
    \frac{\partial h_1(\mu_t)}{\partial \mu_t} & \frac{\partial h_1(\mu_t)}{\partial \sigma^2_t}\\
    \frac{\partial h_2(\sigma^2_t)}{\partial \mu_t} & \frac{\partial h_2(\sigma^2_t)}{\partial \sigma^2_t}
    \end{bmatrix} = \begin{bmatrix} 
    1 & 0\\
    0 & \frac{1}{\sigma^2_t}
    \end{bmatrix}
\end{equation}

Note that $\dot{h}$ is always a diagonal matrix. The reparametrized score derivations for different scalings and different types of maps are presented in Appendix \ref{appendix:parametrizations}. Given the following definitions
\begin{align}
\left.\dot h(f_t) = \frac{\partial h(f_t)}{\partial \tilde f_t}\right|_{f_t}, \qquad \mathcal{J}_{t|t-1} \mathcal{J}_{t|t-1}^{\top} = \mathcal{I}_{t|t-1}^{-1}, \qquad \tilde \nabla_{t} = \left(\dot h\right)^{-1}\nabla_{t},
\end{align}
then the linked scaled score $\tilde{s}_{t}$ can be computed as follows:
\begin{align}
    \tilde s_t &= \left(\dot h\right)^{-1}\nabla_t, \text{ for } d = 0,\label{eq:idscaling} \\
    \tilde s_t &= \mathcal{J}_{t|t-1} \nabla_t, \text{ for } d = \frac{1}{2}, 
    \label{eq:sqrtinvscaling}\\
    \tilde s_t &= \dot h\mathcal{I}_{t|t-1}^{-1}\nabla_t, \text{ for } d = 1.
    \label{eq:invscaling}
\end{align}

\subsection{Maximum Likelihood Estimation}
The vector of hyperparameters $\theta$ can be estimated via maximum likelihood:
\begin{equation}
\hat \theta = {\arg\max}_\theta \sum_{t=1}^N\ln p(y_{t} | f_{t}, \mathcal{F}_{t-1}; \theta)
\end{equation}

Evaluating the log‐likelihood function of the GAS model is particularly simple. Given values for the constant parameters $\theta$, the GAS updating equation \eqref{recursion} outputs the conditional distribution at each time period, which generally has a closed form. Thus, it suffices to look at $\ln p(y_{t} | f_{t}, \mathcal{F}_{t-1}; \theta)$ for a particular value of $\theta$. 

Evaluating the analytical derivatives needed to obtain the maximum likelihood is a demanding and sometimes impossible task. As a consequence, a common practice is to numerically evaluate derivatives using global optimization methods such as \mbox{L-BFGS} \citep{liu1989limited} and Nelder-Mead \citep{neldermead}. Depending on the case, constrained optimization can also be applied, using, for instance, a Newton interior points method. 

\subsection{Forecasting} \label{sec:forecasting}

Forecasting and simulation of future scenarios are among the main goals in time series analysis. In \cite{Blasques2016}, details of the procedure for out-of-sample confidence intervals for the time-varying parameters are discussed. The procedure discussed in Section 4.1 of \cite{Blasques2016} is currently implemented in  \pkg{ScoreDrivenModels.jl} as follows:
\begin{enumerate}
    \item  Given $\hat \theta_T$ and the filtered state $\hat f_{T+1}$, draw $S$ values $y_{T+1}^1, \dots, y_{T+1}^S$ from the estimated conditional density at $T+1$: $y_{T+1} \sim  p(y_{T+1}|\hat f_{T+1}, \hat \theta_T)$ for $s = 1, \dots, S$.
    \item Use $y_{T+1}^1, \dots, y_{T+1}^S$ and the recursion \eqref{recursion} to obtain the filtered values $\hat f_{T+2}^1, \dots, \hat f_{T+2}^S$.
    \item Repeat steps 1 and 2 $H$ times for $H$ steps ahead generating one new value of $y$ and $f$ per scenario $s$.
\end{enumerate}

Once the procedure is over, $S$ scenarios for the observations within the entire horizon, $y_{T+k}^s$ for $k = 1, \dots, H$ and $s=1, \dots,S$ have been simulated. Based on these set of scenario, one can calculate quantile forecasts, build empirical distributions, or use them to feed decision under uncertainty models, such as stochastic programming. Note that this method solely considers the uncertainty of innovations. The consideration of uncertainty on both innovations and parameters, as discussed in \cite{Blasques2016}, is considered future work for the package.

\section{The ScoreDrivenModels.jl Package} \label{sec:package}

\pkg{ScoreDrivenModels.jl} enables users to create and estimate score-driven models and to perform forecasting and simulation while working purely in \proglang{Julia}. Its API allows users to choose between different distributions, scaling values, lag structures, and optimization methods. The basic code structure allows contributors to add new distributions and optimization methods; technical details about adding new features are available in the package documentation. Installation of the package is easily conducted using the Julia Package manager:
\begin{CodeChunk}
\begin{CodeInput}
pkg> add ScoreDrivenModels
\end{CodeInput}
\end{CodeChunk}

\subsection{Model Specification}

To create a \code{Model}, the user must specify 1) the desired distribution, 2) the scaling, 3) the lag structure, and 4) which parameters should be considered time-varying.

\begin{enumerate}
    \item The lag structure in a GAS($p, q$) model can be specified in two ways: either through integers \code{p} and \code{q}, which results in all lags from \code{1} to \code{p} and \code{1} to \code{q} being added, or through arrays of integers \code{ps} and \code{qs} containing only the desired lags.
    \item To specify the distribution, the user needs to choose a distribution among the available ones that have an interface with \pkg{Distributions.jl}. The list of available distributions is displayed in \mbox{Table \ref{tab:distributions}}. Furthermore, we refer the interested reader to Appendix \ref{appendix:scores}, where we provide details on the score calculations for each probability density made available in the package.
    \item The scaling is specified by defining the value of $d$, which can be 0, 1, or $\frac{1}{2}$, respectively the identity scaling, inverse scaling, and inverse square-root scaling.
    \item In order to define which distribution parameters should be time-varying, the keyword argument \code{time_varying_params} can be used. Note that the default behavior is to have all parameters as time-varying.
\end{enumerate}

\begin{table}[h]
\centering
\begin{tabular}{lcccc}
\hline
Distribution           & \begin{tabular}[c]{@{}c@{}}Number of\Tstrut\\parameters\end{tabular} & \begin{tabular}[c]{@{}c@{}}Identity\Tstrut\\scaling\end{tabular} & \begin{tabular}[c]{@{}c@{}}Inverse\Tstrut\\scaling\end{tabular} & \begin{tabular}[c]{@{}c@{}}Inverse\Tstrut\\square-root\\scaling\end{tabular} \\ \hline
Beta & 2 & \checkmark & \checkmark & \checkmark\\
BetaLocationScale & 4 & \checkmark &  -- & --\\
Exponential & 1 & \checkmark & \checkmark & \checkmark\\
Gamma & 2 & \checkmark & \checkmark & \checkmark\\
LogitNormal & 2 & \checkmark & \checkmark & \checkmark\\
LogNormal & 2 & \checkmark & \checkmark & \checkmark\\
NegativeBinomial & 2 & \checkmark & -- & --\\
Normal & 2 & \checkmark & \checkmark & \checkmark\\
Poisson & 1 & \checkmark & \checkmark & \checkmark\\
TDist & 1 & \checkmark & \checkmark & \checkmark\\
TDistLocationScale & 3 & \checkmark & \checkmark & \checkmark\\
Weibull & 2 & \checkmark & -- & --\\
\hline
\end{tabular}
\caption{\label{tab:distributions} List of currently implemented distributions and scalings. $\#$ represent the number of parameters of the distribution.}
\end{table}

Once the model is specified, the unknown parameters that must be estimated are automatically represented as \code{NaN} within the \code{Model} structure. As an example, a GAS($1, 2$) model with lognormal distribution and inverse square-root scaling can be created by writing the following line of code:
\begin{CodeChunk}
\begin{CodeInput}
julia> Model(1, 2, LogNormal, 0.5)
\end{CodeInput}
\begin{CodeOutput}
Model{LogNormal,Float64}([NaN, NaN], Dict(1=>[NaN 0.0; 0.0 NaN]), 
    Dict(2=>[NaN 0.0; 0.0 NaN],1=>[NaN 0.0; 0.0 NaN]), 0.5)
\end{CodeOutput}
\end{CodeChunk}

\code{Dict} is the \proglang{Julia} data structure for dictionaries. Its use allows code flexibility enabling computational simplifications for complex lag structures. As displayed above, the unknown constant parameters to be estimated are set as \code{NaN}. In this case, the constant parameters considered in vector $\omega$ are $A_1$, $B_1$, and $B_2$.

In some applications, however, the user might define only one of the distribution parameters as time-varying. In the example below, the only time-varying parameter is $\mu_{t}$, so the keyword argument \code{time\_varying\_params} indicates a vector with only one element, \code{[1]}, representing the first parameter of the lognormal distribution. A table that indicates the distribution parameters and their orders is available in the package documentation. The choice of the time-varying parameter can be expressed by the following code:

\begin{CodeChunk}
\begin{CodeInput}
julia> Model(1, 2, LogNormal, 0.5; time_varying_params = [1])
\end{CodeInput}
\begin{CodeOutput}
Model{LogNormal,Float64}([NaN, NaN], Dict(1=>[NaN 0.0; 0.0 0.0]),
    Dict(2=>[NaN 0.0; 0.0 0.0],1=>[NaN 0.0; 0.0 0.0]), 0.5)
\end{CodeOutput}
\end{CodeChunk}
Users can also specify the lag structure by passing only the lags of interest. Note that this feature is equivalent to defining that matrices $A_i$ and $B_j$ are equal to zero for certain values $i$ and $j$. An example is a model that uses lags 1 and 12, which means that only the matrices $A_1$, $A_{12}$, $B_1$, and $B_{12}$ have nonzero entries:

\begin{CodeChunk}
\begin{CodeInput}
julia> Model([1, 12], [1, 12], LogNormal, 0.5)
\end{CodeInput}
\begin{CodeOutput}
Model{Normal,Float64}([NaN, NaN], 
    Dict(12=>[NaN 0.0; 0.0 NaN],1=>[NaN 0.0; 0.0 NaN]),
    Dict(12=>[NaN 0.0; 0.0 NaN],1=>[NaN 0.0; 0.0 NaN]), 0.5)
\end{CodeOutput}
\end{CodeChunk}

\subsection{Estimation}

Once the model is specified, the next step is estimation. Users can choose from different optimization methods provided by \pkg{Optim.jl} \citep{mogensen2018optim}.
Since this optimization problem is non-convex, there is no guarantee that the optimal value found by the optimization method is the global optimum. To increase the chances of finding the global optimum, we run the optimization algorithm for different initial parameter values. The default method is Nelder-Mead with 3 random initial parameter values, but the optimization interface is highly flexible. Users can customize convergence tolerances, choose initial parameter values, and, depending on the optimization method, choose bounds for the parameters.
By default, these initial values are the unconditional mean of $f_{t+1}$ which is given by
%
\begin{equation}
    \E[f_{t+1}] = \omega \bigg( I - \sum_{j=1}^{q} B_{j} \bigg)^{-1}. 
\end{equation}

As an illustration, let us estimate a GAS model using the same data and specification used in the R package \pkg{GAS} \citep{Ardia2019} paper with the function \code{fit!}, the data is also available in package repository \citep{bodin2020gas}. The data represents the monthly US inflation measured as the logarithmic change of the consumer price index. The model can be estimated as follows:

\begin{CodeChunk}
\begin{CodeInput}
julia> Random.seed!(123)
julia> y = vec(readdlm("../test/data/cpichg.csv"))
julia> gas = Model(1, 1, TDistLocationScale, 0.0,
                   time_varying_params=[1, 2])
julia> f = fit!(gas, y)
\end{CodeInput}
\begin{CodeOutput}
Round 1 of 3 - Log-likelihood: -178.2064944794775
Round 2 of 3 - Log-likelihood: -178.20649327545632
Round 3 of 3 - Log-likelihood: -178.20649537930336
\end{CodeOutput}
\end{CodeChunk}

Users also have the option to check more detailed results of the optimization procedure by changing the keyword argument \code{verbose}. The default value of this argument is 1; to check the optimization summary, users should set the verbose level to 2, and to see the value of the objective function at each iteration of the optimization, it should be set to 3. To avoid the printing of outputs, users can set \code{verbose = 0}. To illustrate, results with level 2 is depicted below. 

\begin{CodeChunk}
\begin{CodeInput}
julia> gas = Model(1, 1, TDistLocationScale, 0.0,
                   time_varying_params=[1, 2])
julia> f = fit!(gas, y; verbose=2)
\end{CodeInput}
\begin{CodeOutput}
Round 1 of 3 - Log-likelihood: -178.20649402602353
Round 2 of 3 - Log-likelihood: -178.20649463073832
Round 3 of 3 - Log-likelihood: -178.2064932319776

Best initial_point optimization result:
 * Status: success

 * Candidate solution
    Minimizer: [3.74e-02, -2.60e-01, 1.88e+00,  ...]
    Minimum:   1.782065e+02

 * Found with
    Algorithm:     Nelder-Mead
    Initial Point: [3.13e-02, 1.61e-01, 6.53e-02,  ...]

 * Convergence measures
    standard-deviation <= 1.0e-06

 * Work counters
    Seconds run:   0  (vs limit Inf)
    Iterations:    1345
    f(x) calls:    2156
\end{CodeOutput}
\end{CodeChunk}

As mentioned before, while the maximization of the log-likelihood is done by default through the Nelder-Mead method with 3 random initial values, these features can be changed by the user. For example, to use the L-BFGS algorithm with 5 random initial values:

\begin{CodeChunk}
\begin{CodeInput}
julia> gas = Model(1, 1, TDistLocationScale, 0.0,
                   time_varying_params=[1, 2])
julia> f = fit!(gas, y; opt_method=LBFGS(gas, 5))
\end{CodeInput}
\end{CodeChunk}

Once the estimation step is finished, the user can query the results by calling the function \code{fit\_stats}:

\begin{CodeChunk}
\begin{CodeInput}
julia> fit_stats(f)
\end{CodeInput}
\begin{CodeOutput}
--------------------------------------------------------
Distribution:                 Distributions.LocationScale{Float64,
                                           TDist{Float64}}
Number of observations:       276
Number of unknown parameters: 7
Log-likelihood:               -178.2065
AIC:                          370.4130
BIC:                          395.7558
--------------------------------------------------------
Parameter      Estimate   Std.Error     t stat   p-value
omega_1          0.0374      0.0311     1.2016    0.2686
omega_2         -0.2599      0.1409    -1.8454    0.1075
omega_3          1.8758      0.2914     6.4380    0.0004
A_1_11           0.0717      0.0184     3.8884    0.0060
A_1_22           0.4538      0.2139     2.1216    0.0715
B_1_11           0.9432      0.0272    34.6438    0.0000
B_1_22           0.8556      0.0743    11.5141    0.0000
\end{CodeOutput}
\end{CodeChunk}

This result matches the example discussed in \cite{Ardia2019} with the exception of \code{omega_3}, due to a difference in parametrization between the two packages. Once the parameter is recovered to its original parametrization, the result becomes the same.

\subsection{Forecasting and Simulation}

Forecasting in this framework is done by simulation as per Section \ref{sec:forecasting}. Function \code{forecast} runs the procedure proposed by \cite{Blasques2016} and
returns a \code{Forecast} structure that has the expected value for time-varying parameters, observations, and the related scenarios used to find them. By default the structure also stores the 2.5\%, 50\% and 97.5\% quantiles.

Next, we will present forecasting results using the previously estimated US inflation data. In the example below, the first column is the location parameter, the second column is the scale parameter, and the third column represents the degrees of freedom parameter. 

\begin{CodeChunk}
\begin{CodeInput}
julia> forec = forecast(y, gas, 12)
julia> forec.parameter_forecast
\end{CodeInput}
\begin{CodeOutput}
12x3 Array{Float64,2}:
 0.101281  0.152362  6.52618
 0.134314  0.159539  6.52618
 0.165809  0.166048  6.52618
 0.196427  0.171559  6.52618
 0.219774  0.175267  6.52618
 0.245018  0.17878   6.52618
 0.267962  0.179805  6.52618
 0.288717  0.182248  6.52618
 0.308386  0.184336  6.52618
 0.327622  0.185162  6.52618
 0.348588  0.186111  6.52618
 0.36383   0.186892  6.52618
\end{CodeOutput}
\end{CodeChunk}

We can also obtain the scenarios of observations, $y_{T+k}^s$ for $k=1,\dots,H$ and $s=1,\dots,S$, that generated the above forecasted values as follows: 

\begin{CodeChunk}
\begin{CodeInput}
julia> forec.observation_scenarios
\end{CodeInput}
\begin{CodeOutput}
12x10000 Array{Float64,2}:
 1.3839    -0.378146   …   0.561549   0.0415801
 0.971144   0.116446         0.478346   0.290846 
 1.1637     0.439987         0.157433  -0.195271 
 0.50442   -0.0417178        0.204036  -0.549865 
 1.39936    0.149797         0.606875  -0.263888 
 1.04027    0.564609   …  -0.234421   0.139999 
 0.609759   1.2166           0.236292   0.408869 
 0.693315   0.057054         0.802398   0.285412 
 0.908195   1.06543          1.22845    0.256618 
 1.89531    0.566139         0.808426   0.121839 
 1.7671     0.813116   …   0.639728   0.0408979
 0.840164   0.72076          0.316973   0.300578 
\end{CodeOutput}
\end{CodeChunk}

For the sake of clarity, the forecast $\hat{y}_{T+k}$ is the sample average of the scenarios of observations, i.e., $\hat{y}_{T+k}=S^{-1}\sum_{s=1}^S y_{T+k}^s$.

\section{Applications} \label{sec:applications}

\subsection{Hydropower Generation in Brazil}

The Brazilian system operator regularly publishes an indicator of how much energy can be generated from water inflows among all hydropower plants in the country. This indicator is referred to as Affluent Natural Energy (ANE). Usually, ANE is computed in daily basis per region and then aggregated in a monthly basis as the average of each month.

Due to the high dependency of water resources, in Brazil, the monthly ANE is a key component of system operational and infrastructure planning studies. In most decision under uncertainty methodologies, it is essential to have simulated scenarios describing the empirical distribution rather than simple point forecasts. 

\begin{figure}[h!]
\centering
\includegraphics[width=155mm]{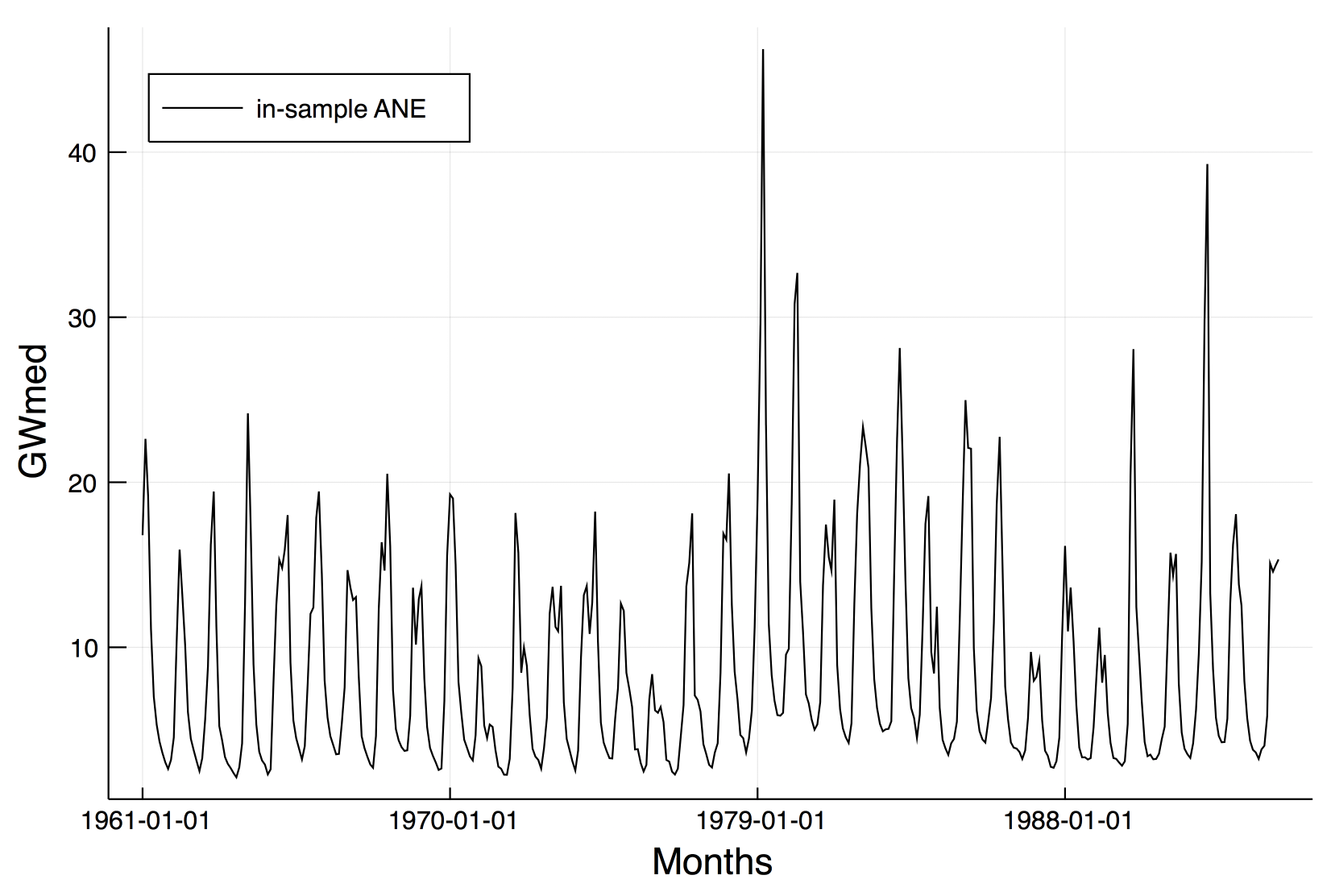}
\caption{\label{fig:ANE} ANE in the Northeastern region of Brazil}
\end{figure}

In this section, we present an example that generates scenarios of the monthly aggregated ANE in the Northeastern region of Brazil illustrated in Fig. \ref{fig:ANE}. This example employs a lognormal GAS model with tailored lag structure, identity scaling and time-varying parameter $\mu_{t}$. Finally, ANE scenarios are simulated based on the fitted hyperparameters.

In this model we utilize lags 1, 2, 11 and 12 for the score and autoregressive components; lags 1 and 2 capture the short-term trend of the series, while lags 11 and 12 capture the seasonal aspect of water inflow. The model can be written as follows:
\begin{equation}
    \left\{\begin{array}{ccl}
    \mu_{t} &=& \omega_1 + A_1s_{t-1} + A_2s_{t-2} + A_{11}s_{t-11} + A_{12}s_{t-12} \\&& +B_1\mu_{t-1} + B_2\mu_{t-2} + B_{11}\mu_{t-11} + B_{12}\mu_{t-12} \\
    \ln(\sigma^2_{t}) &=& \omega_2
    \end{array}
    \label{eq:param_recusion_lognormal} \right.
\end{equation}

\begin{figure}[h!]
\centering
\includegraphics[width=155mm]{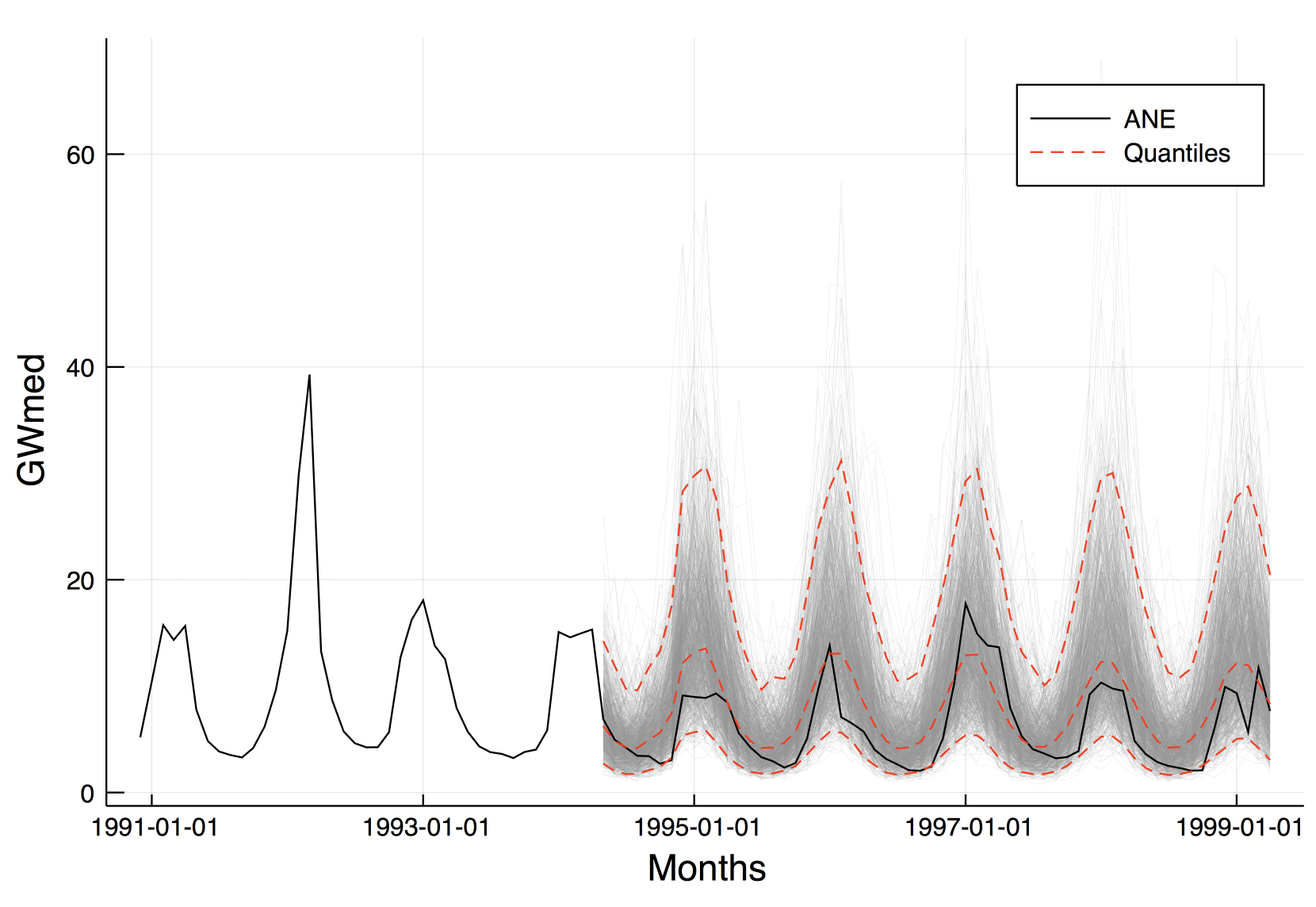}
\caption{\label{fig:ANE_scenarios} ANE scenarios in the Northeastern region of Brazil}
\end{figure}

In this model we have also considered that the initial parameters $\mu_1, \dots \mu_{12}$, $\sigma^2_1, \dots, \sigma^2_{12}$, $s_1,\dots s_{12}$ are calculated heuristically through maximum likelihood estimation in each seasonal component of the model. This way, the parameters $\mu_1, \sigma_1$ are calculated by fitting a lognormal distribution in the observations $y_1, y_{13}, y_{25}, \dots, y_{1 + 12n}$. This procedure was used in \cite{Hoeltgebaum2018}. Note that this procedure, which is implemented by the function \code{dynamic_initial_params} in our package, is relevant for ensuring good estimation results when considering seasonal time series. The model is estimated as follows:

\begin{CodeChunk}
\begin{CodeInput}
julia> Random.seed!(123)
julia> y = vec(readdlm("../test/data/ane_northeastern.csv"))
julia> y_train = y[1:400]
julia> gas = Model([1, 2, 11, 12], [1, 2, 11, 12], LogNormal, 0.0;
+        time_varying_params=[1]
julia> initial_params = dynamic_initial_params(y_train, gas)
julia> f = ScoreDrivenModels.fit!(gas, y_train; initial_params=initial_params)
julia> fit_stats(f)
\end{CodeInput}
\begin{CodeOutput}
--------------------------------------------------------
Distribution:                 LogNormal
Number of observations:       400
Number of unknown parameters: 10
Log-likelihood:               -779.7883
AIC:                          1579.5766
BIC:                          1619.4912
--------------------------------------------------------
Parameter      Estimate   Std.Error     t stat   p-value
omega_1          0.0135      0.0282     0.4807    0.6411
omega_2         -2.8408      0.0720   -39.4651    0.0000
A_1_11          -0.0378      0.0051    -7.4109    0.0000
A_2_11           0.0047      0.0034     1.4052    0.1903
A_11_11         -0.0178      0.0045    -3.9838    0.0026
A_12_11          0.0576      0.0052    11.0468    0.0000
B_1_11          -0.4784      0.0444   -10.7815    0.0000
B_2_11           0.4682      0.0449    10.4210    0.0000
B_11_11         -0.3055      0.0928    -3.2933    0.0081
B_12_11          1.3088      0.0962    13.6080    0.0000
\end{CodeOutput}
\end{CodeChunk}
Once the model is estimated we can generate 1000 ANE scenarios for the next 60 months following the methodology discussed in Section \ref{sec:forecasting}. We will utilize \code{forecast} in order to obtain the quantiles as well. The data set used to estimate the model used data from January 1961 until April 1994. To illustrate the adequacy of our model forecast, we present in Fig. \ref{fig:ANE_scenarios} an out-of-sample study, where the simulated scenarios (gray lines), and related quantiles (red dotted lines) from May 1994 until April 1999 are contrasted to the actual observed values. 
\begin{CodeChunk}
\begin{CodeInput}
julia> forec = forecast(y_train, gas, 60; 
                S = 1_000, initial_params = initial_params)
\end{CodeInput}
\end{CodeChunk}

\subsection{GARCH Model}

One of the advanced features of \pkg{ScoreDrivenModels.jl} is allowing users to change the default parametrization. An example of a different parametrization is the GARCH(1, 1) GAS model. It can be shown that a GAS(1, 1) is equivalent to a GARCH(1, 1) if the function $h$ is the identity (we refer the interested reader to Appendix \ref{diffparam} for further details). To ensure the equivalence, the user must define the identity link function for both time-varying parameters. To do this the user must overwrite three \pkg{ScoreDrivenModel.jl} methods as detailed in the following example:
\begin{CodeChunk}
\begin{CodeInput}
julia> function ScoreDrivenModels.link!(param_tilde::Matrix{T}, 
                            ::Type{Normal}, param::Matrix{T}, t::Int) where T
    param_tilde[t, 1] = link(IdentityLink, param[t, 1])
    param_tilde[t, 2] = link(IdentityLink, param[t, 2])
    return
end
julia> function ScoreDrivenModels.unlink!(param::Matrix{T}, ::Type{Normal}, 
                                   param_tilde::Matrix{T}, t::Int) where T
    param[t, 1] = unlink(IdentityLink, param_tilde[t, 1])
    param[t, 2] = unlink(IdentityLink, param_tilde[t, 2])
    return
end
julia> function ScoreDrivenModels.jacobian_link!(aux::AuxiliaryLinAlg{T}, 
                            ::Type{Normal}, param::Matrix{T}, t::Int) where T
    aux.jac[1] = jacobian_link(IdentityLink, param[t, 1])
    aux.jac[2] = jacobian_link(IdentityLink, param[t, 2])
    return
end
\end{CodeInput}
\end{CodeChunk}

Once the methods have been overwritten for the Normal distribution, the recursion will apply the \code{IdentityLink} in both parameters and the user can proceed to the estimation step. Below, we run an example provided in \cite{broda2020archmodels} for daily German mark/British pound exchange rates. Note that there are also bounds being provided for the hyperparameter estimation through \code{lb} and \code{ub}.
\begin{CodeChunk}
\begin{CodeInput}
julia> y = vec(readdlm("../test/data/BG96.csv"))
julia> initial_params = [mean(y) var(y)]
julia> ub = [1.0, 1.0, 0.5, 1.0]
julia> lb = [-1.0, 0.0, 0.0, 0.5]
julia> gas = Model(1, 1, Normal, 1.0, time_varying_params = [2])
julia> initial_point = [0.0, 0.5, 0.25, 0.75]
julia> f = fit!(gas, y; initial_params = initial_params,
                       opt_method = IPNewton(gas, [initial_point] 
                       ub=ub, lb=lb))
Round 1 of 1 - Log-likelihood: -1106.598367006442
julia> fit_stats(f)
\end{CodeInput}
\begin{CodeOutput}
--------------------------------------------------------
Distribution:                 Normal
Number of observations:       1974
Number of unknown parameters: 4
Log-likelihood:               -1106.5984
AIC:                          2221.1967
BIC:                          2243.5480
--------------------------------------------------------
Parameter      Estimate   Std.Error     t stat   p-value
omega_1         -0.0062      0.0085    -0.7373    0.5019
omega_2          0.0108      0.0029     3.7732    0.0196
A_1_22           0.1534      0.0266     5.7726    0.0045
B_1_22           0.9593      0.0144    66.6015    0.0000
\end{CodeOutput}
\end{CodeChunk}

The results obtained above are the same as the ones found in the Section 4.1 of \cite{broda2020archmodels}, with the exception of one parameter. This is due to a difference in the parametrizations of the models -- the demonstration of the equivalence between the hyperparameters of the GAS model and the GARCH model can be found in Appendix \ref{diffparam}. Note that, following the demonstration in \ref{diffparam}, we have $\beta_1 = B_1 - A_1 = 0.8059$, which is exactly the reported value in \cite{broda2020archmodels}.

We have also run the same example using the R package \pkg{rugarch} \citep{rugarch} and obtained the same result, thus further illustrating the equivalence of the GARCH(1, 1) and the Normal GAS(1, 1) under this parametrization.

\newpage

\section{Conclusion} \label{sec:conclusion}

The \pkg{ScoreDrivenModels.jl} package provides a general framework for score-driven models and represents a user-friendly, off-the-shelf tool. The package is fully implemented in Julia, thus not depending on subroutines written in lower-level languages such as \proglang{C} or \proglang{Fortran}. The model specification is flexible and allows defining any desired lag structure, as well as any distribution due to a tailored dependency on the \pkg{Distributions.jl} package. The estimation procedure is based on numerical optimization algorithms such as Nelder-Mead or L-BFGS and employs the well-known package \pkg{Optim.jl}. Special initialization procedures are implemented to robustify the estimation process for the case of seasonal time series. Available forecasting and simulation procedures allow users to study future data from the estimated model. Finally, the examples provided in Section \ref{sec:applications} illustrate the functionalities of the package as well as possible applications. The software continues to evolve, new features such as an heuristic for the initial hyperparameter and unobserved components modeling are considered as future research topics. The software documentation can be found in \url{https://lampspuc.github.io/ScoreDrivenModels.jl/latest/}


\section*{Acknowledgements}

This study was funded in part by the Coordenação de Aperfeiçoamento de Pessoal de Nível Superior - Brasil (CAPES) - Finance Code 001, by the Conselho Nacional de Desenvolvimento Científico e Tecnológico (CNPq), and by the Energisa Group through the R\&D project ANEEL PD-00405-1701/2017.

\bibliography{refs}

\begin{thebibliography}{26}
\newcommand{\enquote}[1]{``#1''}
\providecommand{\natexlab}[1]{#1}
\providecommand{\url}[1]{\texttt{#1}}
\providecommand{\urlprefix}{URL }
\expandafter\ifx\csname urlstyle\endcsname\relax
  \providecommand{\doi}[1]{doi:\discretionary{}{}{}#1}\else
  \providecommand{\doi}{doi:\discretionary{}{}{}\begingroup
  \urlstyle{rm}\Url}\fi
\providecommand{\eprint}[2][]{\url{#2}}

\bibitem[{Ardia \emph{et~al.}(2019)Ardia, Boudt, and Catania}]{Ardia2019}
Ardia D, Boudt K, Catania L (2019).
\newblock \enquote{{Generalized autoregressive score models in R: The GAS
  package}.}
\newblock \emph{Journal of Statistical Software}, \textbf{88}(1).
\newblock ISSN 15487660.
\newblock \doi{10.18637/jss.v088.i06}.
\newblock \eprint{1609.02354}.

\bibitem[{Ayala and Blazsek(2018)}]{ayala2018score}
Ayala A, Blazsek S (2018).
\newblock \enquote{Score-driven copula models for portfolios of two risky
  assets.}
\newblock \emph{The European Journal of Finance}, \textbf{24}(18), 1861--1884.
\newblock \doi{10.1080/1351847X.2018.1464488}.

\bibitem[{Bezanson \emph{et~al.}(2017)Bezanson, Edelman, Karpinski, and
  Shah}]{bezanson2017julia}
Bezanson J, Edelman A, Karpinski S, Shah V (2017).
\newblock \enquote{Julia: A fresh approach to numerical computing.}
\newblock \emph{SIAM review}, \textbf{59}(1), 65--98.
\newblock \doi{10.1137/141000671}.

\bibitem[{Blasques \emph{et~al.}(2016)Blasques, Koopman, {\L}asak, and
  Lucas}]{Blasques2016}
Blasques F, Koopman SJ, {\L}asak K, Lucas A (2016).
\newblock \enquote{{In-sample confidence bands and out-of-sample forecast bands
  for time-varying parameters in observation-driven models}.}
\newblock \emph{International Journal of Forecasting}, \textbf{32}(3),
  875--887.
\newblock \doi{10.1016/j.ijforecast.2015.11.018}.

\bibitem[{Bodin \emph{et~al.}(2020)Bodin, Saavedra, and
  Fernandes}]{bodin2020gas}
Bodin G, Saavedra R, Fernandes C (2020).
\newblock \enquote{{\texttt{ScoreDrivenModels.jl} [Online]. Available:
  \url{https://github.com/LAMPSPUC/ScoreDrivenModels.jl}}.}

\bibitem[{Bollerslev(1986)}]{bollerslev1986generalized}
Bollerslev T (1986).
\newblock \enquote{Generalized autoregressive conditional heteroskedasticity.}
\newblock \emph{Journal of econometrics}, \textbf{31}(3), 307--327.

\bibitem[{Broda and Paolella(2020)}]{broda2020archmodels}
Broda SA, Paolella MS (2020).
\newblock \enquote{{ARCHModels.jl: Estimating ARCH models in Julia}.}

\bibitem[{Cox \emph{et~al.}(1981)Cox, Gudmundsson, Lindgren, Bondesson,
  Harsaae, Laake, Juselius, and Lauritzen}]{cox1981parameterdriven}
Cox DR, Gudmundsson G, Lindgren G, Bondesson L, Harsaae E, Laake P, Juselius K,
  Lauritzen SL (1981).
\newblock \enquote{Statistical Analysis of Time Series: Some Recent
  Developments [with Discussion and Reply].}
\newblock \emph{Scandinavian Journal of Statistics}, \textbf{8}(2), 93--115.
\newblock ISSN 03036898, 14679469.
\newblock \urlprefix\url{http://www.jstor.org/stable/4615819}.

\bibitem[{Creal \emph{et~al.}(2013)Creal, Koopman, and
  Lucas}]{creal2013generalized}
Creal D, Koopman SJ, Lucas A (2013).
\newblock \enquote{{Generalized autoregressive score models with
  applications}.}
\newblock \emph{Journal of Applied Econometrics}, \textbf{28}(5), 777--795.
\newblock ISSN 08837252.
\newblock \doi{10.1002/jae.1279}.

\bibitem[{de~Melo \emph{et~al.}(2018)de~Melo, Fernandes, and
  de~Melo}]{demelo2018forecasting}
de~Melo MAB, Fernandes CA, de~Melo EF (2018).
\newblock \enquote{Forecasting aggregate claims using score-driven time series
  models.}
\newblock \emph{Statistica Neerlandica}, \textbf{72}(3), 354--374.
\newblock \doi{10.1111/stan.12139}.

\bibitem[{Durbin and Koopman(2012)}]{durbin2012time}
Durbin J, Koopman SJ (2012).
\newblock \emph{Time Series Analysis by State Space Methods}.
\newblock Oxford University Press.
\newblock \doi{10.1093/acprof:oso/9780199641178.001.0001}.

\bibitem[{Engle and Russell(1998)}]{engle1998duration}
Engle RF, Russell JR (1998).
\newblock \enquote{Autoregressive Conditional Duration: A New Model for
  Irregularly Spaced Transaction Data.}
\newblock \emph{Econometrica}, \textbf{66}(5), 1127--1162.
\newblock ISSN 00129682, 14680262.
\newblock \urlprefix\url{http://www.jstor.org/stable/2999632}.

\bibitem[{Ghalanos(2020)}]{rugarch}
Ghalanos A (2020).
\newblock \enquote{{rugarch: Univariate GARCH Models. R package version 1.4-2
  [Online]. Available: \url{https://CRAN.R-project.org/package=rugarch}}.}

\bibitem[{Harvey(2013)}]{harvey2013dynamic}
Harvey AC (2013).
\newblock \emph{Dynamic Models for Volatility and Heavy Tails: With
  Applications to Financial and Economic Time Series}.
\newblock Econometric Society Monographs. Cambridge University Press.
\newblock \doi{10.1017/CBO9781139540933}.

\bibitem[{Harvey and Thiele(2016)}]{harvey2016testing}
Harvey AC, Thiele S (2016).
\newblock \enquote{Testing against changing correlation.}
\newblock \emph{Journal of Empirical Finance}, \textbf{38}(Part B), 575--589.
\newblock ISSN 09275398.
\newblock \doi{10.1016/j.jempfin.2015.09.003}.

\bibitem[{Hoeltgebaum \emph{et~al.}(2018)Hoeltgebaum, Fernandes, and
  Street}]{Hoeltgebaum2018}
Hoeltgebaum H, Fernandes C, Street A (2018).
\newblock \enquote{{Generating joint scenarios for renewable generation: The
  case for non-gaussian models with time-varying parameters}.}
\newblock \emph{IEEE Transactions on Power Systems}, \textbf{33}(6),
  7011--7019.
\newblock ISSN 08858950.
\newblock \doi{10.1109/TPWRS.2018.2838050}.

\bibitem[{Koopman \emph{et~al.}(2000)Koopman, Harvey, Doornik, and
  Shephard}]{koopman2000stamp}
Koopman SJ, Harvey AC, Doornik JA, Shephard N (2000).
\newblock \enquote{{STAMP} 6.0: Structural Time Series Analyser, Modeller and
  Predictor.}
\newblock \emph{London: Timberlake Consultants}.

\bibitem[{Liu and Nocedal(1989)}]{liu1989limited}
Liu DC, Nocedal J (1989).
\newblock \enquote{On the limited memory {BFGS} method for large scale
  optimization.}
\newblock \emph{Mathematical programming}, \textbf{45}(1-3), 503--528.

\bibitem[{Mogensen and Riseth(2018)}]{mogensen2018optim}
Mogensen PK, Riseth AN (2018).
\newblock \enquote{Optim: A mathematical optimization package for {J}ulia.}
\newblock \emph{Journal of Open Source Software}, \textbf{3}(24).
\newblock \doi{10.21105/joss.00615}.

\bibitem[{Nani \emph{et~al.}(2019)Nani, Gamoudi, and
  El~Ghourabi}]{nani2019value}
Nani A, Gamoudi I, El~Ghourabi M (2019).
\newblock \enquote{Value-at-risk estimation by LS-SVR and FS-LS-SVR based on
  GAS model.}
\newblock \emph{Journal of Applied Statistics}, \textbf{46}(12), 2237--2253.
\newblock \doi{10.1080/02664763.2019.1584161}.

\bibitem[{Nelder and Mead(1965)}]{neldermead}
Nelder JA, Mead R (1965).
\newblock \enquote{{A Simplex Method for Function Minimization}.}
\newblock \emph{The Computer Journal}, \textbf{7}(4), 308--313.
\newblock ISSN 0010-4620.
\newblock \doi{10.1093/comjnl/7.4.308}.
\newblock
  \eprint{https://academic.oup.com/comjnl/article-pdf/7/4/308/1013182/7-4-308.pdf},
  \urlprefix\url{https://doi.org/10.1093/comjnl/7.4.308}.

\bibitem[{Neves \emph{et~al.}(2017)Neves, Fernandes, and
  Hoeltgebaum}]{Neves2017}
Neves C, Fernandes C, Hoeltgebaum H (2017).
\newblock \enquote{{Five different distributions for the Lee-Carter model of
  mortality forecasting: A comparison using GAS models}.}
\newblock \emph{Insurance: Mathematics and Economics}, \textbf{75}, 48--57.
\newblock ISSN 01676687.
\newblock \doi{10.1016/j.insmatheco.2017.04.004}.

\bibitem[{Patton \emph{et~al.}(2019)Patton, Ziegel, and
  Chen}]{patton2019dynamic}
Patton AJ, Ziegel JF, Chen R (2019).
\newblock \enquote{Dynamic semiparametric models for expected shortfall (and
  value-at-risk).}
\newblock \emph{Journal of Econometrics}, \textbf{211}(2), 388--413.

\bibitem[{Saavedra(2017)}]{saavedra2017study}
Saavedra R (2017).
\newblock \enquote{A study on the impact of {El Niño Southern Oscillation} on
  hydro power generation in {Brazil}.}
\newblock \doi{10.17771/PUCRio.acad.32290}.

\bibitem[{Saavedra \emph{et~al.}(2019)Saavedra, Bodin, and
  Souto}]{saavedra2019statespacemodels}
Saavedra R, Bodin G, Souto M (2019).
\newblock \enquote{StateSpaceModels.jl: a Julia Package for Time-Series
  Analysis in a State-Space Framework.}
\newblock \emph{arXiv preprint arXiv:1908.01757}.

\bibitem[{Taylor(2016)}]{pyfluxTaylor}
Taylor R (2016).
\newblock \enquote{{PyFlux}.}
\newblock [Online]. Available: \url{https://github.com/RJT1990/pyflux}.

\end{thebibliography}

\begin{appendix}

\section{Parametrizations}\label{appendix:parametrizations}

The use of different link functions and scaling coefficient give rise to different probability function parametrizations and models. Thus, in this appendix we explore some modeling variants due to the choice of link functions and values for the scaling coefficient. 

\subsection{Possible Parametrizations}

There are three mapping functions available in \pkg{ScoreDrivenModels.jl}. Within the context of the software we call them \code{Links}. Each one of them can be used as the default mapping function for a given distribution implemented in the package.
\begin{itemize}
    \item Identity link: $\tilde{f} = f \text{ where } f \in \mathbb{R} \text{ and } \tilde{f} \in \mathbb{R}$
    \item Log link: $\tilde{f} = \ln \left(f - a\right) \text{ where } f \in [a, \infty) \text{ and } \tilde{f} \in \mathbb{R}$
    \item Logit link: $\tilde{f} = \ln\left(\frac{f - a}{b - f}\right) \text{ where } f \in [a, b] \text{ and } \tilde{f} \in \mathbb{R}$
\end{itemize}

\subsection{Score Derivations for Different Scaling Values}

In this subsection we derive expressions \eqref{eq:idscaling}, \eqref{eq:sqrtinvscaling}, and \eqref{eq:invscaling}.

\subsubsection{Scaling d = 0}
For $d = 0$ the score is simply equal to
\begin{equation}
    \tilde \nabla_{t} = \frac{\partial \ln p(y_t|y_{t-1}, f_t)}{\partial \tilde f_t},
\end{equation}
which is equivalent to
\begin{equation}
    \tilde \nabla_{t} = \frac{\partial f_t}{\partial \tilde f_t} \cdot \frac{\partial \ln p(y_t|y_{t-1}, f_t)}{\partial f_t}.
\end{equation}
Notice that one can show by the inverse function theorem that this is the inverse of the Jacobian $\dot h$. Thus, it follows that
\begin{equation}
    \tilde \nabla_{t} = \left(\dot h\right)^{-1}\nabla_t.
\end{equation}
By proceeding the calculus above indicated, we show that
\begin{align}
    \tilde \nabla_t = \begin{bmatrix} 
    \frac{\partial \ln p(y_t|y_{t-1}, \mu, \sigma^2)}{\partial \mu}\\
    \frac{\partial \ln p(y_t|y_{t-1}, \mu, \sigma^2)}{\partial \tilde \sigma^2}\\
    \end{bmatrix} &=
    \left(\begin{bmatrix} 
    1 & 0\\
    0 & \frac{1}{\sigma^2_t}
    \end{bmatrix}\right)^{-1}
    \begin{bmatrix} 
    \frac{\partial \ln p(y_t|y_{t-1}, \mu, \sigma^2)}{\partial \mu}\\
    \frac{\partial \ln p(y_t|y_{t-1}, \mu, \sigma^2)}{\partial \sigma^2}\\
    \end{bmatrix}\\
    \tilde \nabla_t = \begin{bmatrix} 
    \frac{\partial \ln p(y_t|y_{t-1}, \mu, \sigma^2)}{\partial \tilde \mu}\\
    \frac{\partial \ln p(y_t|y_{t-1}, \mu, \sigma^2)}{\partial \tilde \sigma^2}\\
    \end{bmatrix} &=
    \begin{bmatrix} 
    1 & 0\\
    0 & \sigma^2_t
    \end{bmatrix}
    \begin{bmatrix} 
    \frac{\partial \ln p(y_t|y_{t-1}, \mu, \sigma^2)}{\partial \mu}\\
    \frac{\partial \ln p(y_t|y_{t-1}, \mu, \sigma^2)}{\partial \sigma^2}
    \end{bmatrix}\\
    \tilde \nabla_t = \begin{bmatrix} 
    \frac{\partial \ln p(y_t|y_{t-1}, \mu, \sigma^2)}{\partial \tilde \mu}\\
    \frac{\partial \ln p(y_t|y_{t-1}, \mu, \sigma^2)}{\partial \tilde \sigma^2}\\
    \end{bmatrix} &=
    \begin{bmatrix} 
    \frac{\partial \ln p(y_t|y_{t-1}, \mu, \sigma^2)}{\partial \mu}\\
    \frac{\partial \ln p(y_t|y_{t-1}, \mu, \sigma^2)}{\partial \sigma^2} \cdot \sigma^2_t
    \end{bmatrix}.
\end{align}

\subsubsection{Scaling d = 1/2}
The original scaled score is $s_t = \mathcal{J}_{t|t-1}\nabla_t$ as in considered in \cite{creal2013generalized}. Thus, recall that
\begin{equation}
    \mathcal{I}_{t|t-1}^{-1} = \mathcal{J}_{t|t-1}\mathcal{J}_{t|t-1}^\top.
\end{equation}
In this case, the new scaled score is $\tilde s_t = \mathcal{\tilde J}_{t|t-1}\tilde \nabla_t$. Thus, if $\tilde \nabla_t = \left(\dot h\right)^{-1}\nabla_t$ and $\mathcal{\tilde J}_{t|t-1}$ is yet to be calculated, we have
\begin{align}
    \mathcal{I}_{t|t-1} &= E\left[\nabla_t\nabla_t^\top\right]\\
    \mathcal{\tilde I}_{t|t-1} &= E\left[\left(\dot h\right)^{-1}\nabla_t\nabla_t^\top\left(\dot h\right)^{-1}\right]\\
    \mathcal{\tilde I}_{t|t-1} &= \left(\dot h\right)^{-1}E\left[\nabla_t\nabla_t^\top\right]\left(\dot h\right)^{-1}\\
    \mathcal{\tilde I}_{t|t-1} &= \left(\dot h\right)^{-1}\mathcal{I}_{t|t-1}\left(\dot h\right)^{-1}.
\end{align}
As $\left(\dot h\right)^{-1}$ is a diagonal matrix, in the above development we omitted the transpose operator. So, the inverse of the reparametrized information matrix is equal to
\begin{align}
    \mathcal{\tilde I}_{t|t-1}^{-1} &= \dot h\mathcal{I}_{t|t-1}^{-1}\dot h\\
    \mathcal{\tilde I}_{t|t-1}^{-1} &= \dot h\mathcal{J}_{t|t-1}\mathcal{J}_{t|t-1}^\top\dot h.
\end{align}
Hence, it follows that
\begin{align}
    \mathcal{\tilde J}_{t|t-1}\mathcal{\tilde J}_{t|t-1}^\top &= \dot h\mathcal{J}_{t|t-1}\mathcal{J}_{t|t-1}^\top\dot h\\
    \mathcal{\tilde J}_{t|t-1} &= \dot h\mathcal{J}_{t|t-1},
\end{align}
which lead us to conclude that for this type of scaling $\tilde s_t = s_t$. This becomes clear in the following development: 
\begin{align}
    \tilde s_t &= \mathcal{\tilde J}_{t|t-1}\tilde \nabla_t\\
    \tilde s_t &= \dot h\mathcal{J}_{t|t-1} \left(\dot h\right)^{-1}\nabla_t,
\end{align}
In the case that $\mathcal{J}_{t|t-1}$ is diagonal, because $\dot h$ is diagonal we have 
\begin{equation}
    \tilde s_t = \mathcal{J}_{t|t-1} \nabla_t = s_t.
\end{equation}

\subsubsection{Scaling d = 1}
In this case, the scaled score is equal to $s_t = \mathcal{I}_{t|t-1}^{-1} \nabla_t$ and the reparametrized scaled score is equal to $\tilde s_t = \mathcal{\tilde I}_{t|t-1}^{-1} \tilde \nabla_t$. As previously calculated we have that
\begin{align}
    \mathcal{\tilde I}_{t|t-1}^{-1} &= \dot h\mathcal{I}_{t|t-1}^{-1}\dot h\\
     \tilde \nabla_{t} &= \left(\dot h\right)^{-1}\nabla_t.
\end{align}
Therefore,  
\begin{align}
    \tilde s_t &= \mathcal{\tilde I}_{t|t-1}^{-1}\tilde \nabla_t\\
     \tilde s_t &= \dot h\mathcal{I}_{t|t-1}^{-1}\dot h \left(\dot h\right)^{-1}\nabla_t\\
     \tilde s_t &= \dot h\mathcal{I}_{t|t-1}^{-1}\nabla_t\\
     \tilde s_t &= \dot h s_t.
\end{align}

\subsection{Different Parametrizations Lead to Different Models}\label{diffparam}

One of the examples given in \cite{creal2013generalized} shows the equivalence between a GARCH(1, 1) and GAS(1, 1) with Normal distribution and $d = 1$. An important note on this fact is that the models are only equivalent if the variance, $\sigma^2$, is considered a time-varying parameter, i.e., if the link is the identity function. Therefore, if a log link is used to ensure a positive value for $\sigma^2$, for instance, the equivalence will not hold. As this calculations are not shown in any other work and they reveal relevant insights that can be tested in through our software, we provide further details in the sequel. 

Thus, let us develop the GAS recursion for both cases. Recall the Normal probability density function (pdf):
\begin{equation}
    p(y_t|y_{t-1}, \mu, \sigma^2) = \frac{1}{\sqrt{2\pi\sigma^2}}e^{ \left({\frac{-(y_t - \mu)^2}{2\sigma^2}}\right)}.
\end{equation}
The log-pdf is:
\begin{equation}
    \ln p(y_t|y_{t-1}, \mu, \sigma^2) = \frac{-1}{2}\ln 2 \pi -\frac{1}{2}\ln \sigma^2 - \frac{1}{2}\frac{\left(y_t -\mu \right)^2}{\sigma^2}.
\end{equation}
To calculate the score with respect to $\sigma^2$ we need to calculate the following derivative:
\begin{equation}
    \frac{\partial \ln p(y_t|y_{t-1}, \mu, \sigma^2)}{\partial \sigma^2} = \frac{-1}{2\sigma^2} +  \frac{(y_t - \mu)^2}{2\sigma^4}.
\end{equation}
Then, we calculate the Fisher information as follows:
\begin{equation}
    -E\left[\frac{\partial^2 \ln p(y_t|y_{t-1}, \mu, \sigma^2)}{\partial \sigma^2\partial \sigma^2}\right] = \frac{1}{2\sigma^4}.
\end{equation}
Now if we proceed to write the GAS(1,1) recursion using the inverse scaling $d = 1$, we find
\begin{align}
    \sigma^2_{t+1} &= \omega + A_{1}s_{t} + B_{1}\sigma^2_{t}\\
    \sigma^2_{t+1} &= \omega + A_{1}((y_t - \mu_t)^2 - \sigma^2_{t}) + B_{1}\sigma^2_{t}.
\end{align}
By assuming $\mu = 0$ under correct specification, the recursion becomes
\begin{equation}
    \sigma^2_{t+1} = \omega + A_{1}(y_t^2 - \sigma^2_{t}) + B_{1}\sigma^2_{t}
    \label{eq:gas_1_1},
\end{equation}
which is equivalent to the GARCH(1,1) model
\begin{equation}
    \sigma^2_{t+1} = \alpha_0 + \alpha_1y_t^2 + \beta_{1}\sigma^2_{t}
    \label{eq:garch_1_1}.
\end{equation}
Note that there is sufficient degrees of freedom to make the correspondence between the parameters of the two models, e.g., $\alpha_0 = \omega, \alpha_1 = A_1, \beta_1 = B_1 - A_1$.

Now let us work with 
a different parametrization to assure a positive value to $\sigma^2$. The approach suggested in \cite{creal2013generalized} is to use a map $h(\cdot)=\ln(\cdot)$, i.e., $\tilde \sigma^2 = \ln \sigma^2$. When we use this parametrization the recursion adapts the following way:
\begin{equation}\left\{\begin{array}{ccl}
    \sigma^2_{t} &=& h^{-1}(\tilde \sigma^2_t), \\
    \tilde \sigma^2_{t+1} &=& \omega + A_{1}\tilde s_t + B_{1} \tilde \sigma^2_{t}.
    \end{array}
    \right.
\end{equation}
And we shown in the previous section that $\tilde s_t = \dot h s_t$. In this case $\dot h = \frac{1}{\sigma^2_t}$, so the recursion assumes the following form:
\begin{equation}\left\{\begin{array}{ccl}
    \sigma^2_{t} &=& h^{-1}(\tilde \sigma^2_t), \\
    \tilde \sigma^2_{t+1} &=& \omega + A_{1}\frac{y_t^2 - \sigma^2_t}{\sigma^2_t} + B_{1} \tilde \sigma^2_{t}.
    \end{array}
    \right.
\end{equation}
If we rewrite the recursion solely in terms of $\sigma^2$ we have
\begin{equation}
    \ln \left(\sigma^2_{t+1}\right) = \omega + A_{1}\frac{y_t^2 - \sigma^2_t}{\sigma^2_t} + B_{1} \ln\left(\sigma^2_{t}\right).
    \label{eq:gas_para_1_1}
\end{equation}
In this case however, it is impossible to choose the parameters $\omega, A_1$ and $B_1$ to meet a recursion equivalent to \eqref{eq:garch_1_1}.

\section{Score Calculations}\label{appendix:scores}

GAS models can be still considered a relatively recent technology. In this context, this paper also aims to provide users with a technical reference for the \pkg{ScoreDrivenModels.jl} software. Therefore, in this Appendix, we provide detailed information on the score calculation for each distribution considered in the package.

\subsection{Beta}
Density function
\begin{align*}
    &p(y_t|y_{t-1}, \alpha, \beta) = \frac{y_{t}^{\alpha - 1}(1-y_t)^{\beta - 1}}{B(\alpha, \beta)} \quad \text{where} \quad B(\alpha, \beta) = \frac{\Gamma(\alpha)\Gamma(\beta)}{\Gamma(\alpha + \beta)},\\ & \qquad \qquad \qquad \qquad \qquad \qquad \qquad \qquad \alpha \in \mathbb{R}^+, \quad \beta \in \mathbb{R}^+, \quad \Gamma(\cdot) \text{ is the Gamma function.} \\
    &\E[y_t|y_{t-1}] = \frac{\alpha}{\alpha+\beta},\\ &\VAR[y_t|y_{t-1}] = \frac{\alpha\beta}{(\alpha+\beta)^2(\alpha+\beta+1)}
\end{align*}
Score calculation
\begin{align*}
\nabla_{t} &= \begin{bmatrix} 
\frac{\partial \ln p(y_t|y_{t-1}, \alpha, \beta)}{\partial \alpha}\\
\frac{\partial \ln p(y_t|y_{t-1}, \alpha, \beta)}{\partial \beta}\\
\end{bmatrix}\\
\ln p(y_t|y_{t-1},a, c, \alpha, \beta) &= \left(\alpha-1\right) \ln y_t + \left(\beta-1\right) \ln (1-y_t) - \ln B(\alpha, \beta) \\
\nabla_{t}^{\alpha} = \frac{\partial \ln p(y_t|y_{t-1}, \alpha, \beta)}{\partial \alpha} &= \ln y_t + \psi(\alpha+\beta) - \psi(\alpha)\\
\nabla_{t}^{\beta} = \frac{\partial \ln p(y_t|y_{t-1},a, c, \alpha, \beta)}{\partial \beta} &= \ln (1-y_t) + \psi(\alpha+\beta) - \psi(\beta)
\end{align*}

\subsection{Beta location scale}
Density function
\begin{align*}
    &p(y_t|y_{t-1},a, c, \alpha, \beta) = \frac{\left(y_{t} - a\right)^{\alpha - 1}(c-y_t)^{\beta - 1}}{\left(c - a\right)B(\alpha, \beta)} \quad \text{where} \quad B(\alpha, \beta) = \frac{\Gamma(\alpha)\Gamma(\beta)}{\Gamma(\alpha + \beta)}, \\
    & \qquad \qquad \qquad \qquad \qquad \qquad \qquad \qquad \qquad \qquad \qquad  a, c \in \mathbb{R}
    \quad \alpha, \beta \in \mathbb{R}^+\\
    &\E[y_t|y_{t-1}] = a + \left(c-a\right)\frac{\alpha}{\alpha+\beta},\\ &\VAR[y_t|y_{t-1}] = \left(c-a\right)^2\frac{\alpha\beta}{(\alpha+\beta)^2(\alpha+\beta+1)}
\end{align*}
Score calculation
\begin{align*}
\nabla_{t} &= \begin{bmatrix} 
\frac{\partial \ln p(y_t|y_{t-1},a,c, \alpha, \beta)}{\partial a}\\
\frac{\partial \ln p(y_t|y_{t-1},a,c, \alpha, \beta)}{\partial c}\\
\frac{\partial \ln p(y_t|y_{t-1},a,c, \alpha, \beta)}{\partial \alpha}\\
\frac{\partial \ln p(y_t|y_{t-1},a,c, \alpha, \beta)}{\partial \beta}\\
\end{bmatrix}\\
\ln p(y_t|y_{t-1},a, c, \alpha, \beta) &= \left(\alpha-1\right) \ln \left(y_t - a\right) + \left(\beta-1\right) \ln (c-y_t) \\
& -\left(\alpha + \beta - 1\right)\ln\left(c-a\right) - \ln B(\alpha, \beta) \\
\nabla_{t}^{a} = \frac{\partial \ln p(y_t|y_{t-1},a,c, \alpha, \beta)}{\partial a} &= \frac{- \alpha + 1}{y - a} + \frac{\alpha + \beta - 1}{c - a}\\
\nabla_{t}^{c} = \frac{\partial \ln p(y_t|y_{t-1},a,c, \alpha, \beta)}{\partial c} &= \frac{\beta - 1}{c - y} - \frac{\alpha + \beta - 1}{c - a}\\
\nabla_{t}^{\alpha} = \frac{\partial \ln p(y_t|y_{t-1},a,c, \alpha, \beta)}{\partial \alpha} &= \ln (y_t - a) - \ln (c - a) + \psi(\alpha+\beta) - \psi(\alpha)\\
\nabla_{t}^{\beta} = \frac{\partial \ln p(y_t|y_{t-1},a,c, \alpha, \beta)}{\partial \beta} &= \ln (c-y_t) - \ln (c - a) + \psi(\alpha+\beta) - \psi(\beta)
\end{align*}

\subsection{Exponential}
Density function
\begin{align*}
    &p(y_t|y_{t-1},\lambda) = \lambda e^{-\lambda y_t}
 \quad \lambda \in \mathbb{R}^+\\
    &\E[y_t|y_{t-1}] = \frac{1}{\lambda},\\ &\VAR[y_t|y_{t-1}] = \frac{1}{\lambda^2}
\end{align*}
Score calculation
\begin{align*}
\nabla_{t} &= \begin{bmatrix} 
\frac{\partial \ln p(y_t|y_{t-1}, \lambda)}{\partial \lambda}
\end{bmatrix}\\
\ln p(y_t|y_{t-1}, \lambda, k) &= \ln \lambda - e^{-\lambda y_t}\\
\nabla_{t}^{\lambda} = \frac{\partial \ln p(y_t|y_{t-1}, \lambda)}{\partial \lambda} &= \frac{1}{\lambda} - y_t
\end{align*}

\subsection{Gamma}
Density function
\begin{align*}
    &p(y_t|y_{t-1}, \alpha, k) ={\frac{y_t^{\alpha-1}e^{-{\frac{y_t}{k}}}}{\Gamma (\alpha)k ^{\alpha}}}, \quad \alpha \in \mathbb{R}^+ , \quad k \in \mathbb{R}^+\\
    &\E[y_t|y_{t-1}] = \alpha k,\\ &\VAR[y_t|y_{t-1}] = \alpha k^2
\end{align*}
Score calculation
\begin{align*}
\nabla_{t} &= \begin{bmatrix} 
\frac{\partial \ln p(y_t|y_{t-1}, \alpha, k)}{\partial \alpha}\\
\frac{\partial \ln p(y_t|y_{t-1}, \alpha, k)}{\partial k}\\
\end{bmatrix}\\
\ln p(y_t|y_{t-1}, \alpha, k) &= (\alpha - 1)\ln y_t - \frac{y_t}{k} - \ln\Gamma(\alpha) - \alpha \ln k \\
\nabla_{t}^{\alpha} = \frac{\partial \ln p(y_t|y_{t-1}, \alpha, k)}{\partial \alpha} &= \ln y_t - \psi(\alpha) - \ln k\\
\nabla_{t}^{k} = \frac{\partial \ln p(y_t|y_{t-1}, \alpha, k)}{\partial k} &= \frac{y_t}{k^2} - \frac{\alpha}{k}
\end{align*}

\subsection{Logit-Normal}
Density function
\begin{align*}
    &p(y_t|y_{t-1}, \mu, \sigma^2) = \frac{1}{y_t(1 - y_t){\sqrt {2\pi\sigma^2}}}\ e^{\left(-{\frac {\left(\text{logit}\left(y_t\right)-\mu \right)^{2}}{2\sigma ^{2}}}\right)} , \quad \mu \in \mathbb{R}, \quad \sigma^2 \in \mathbb{R}^+\\
    &\E[y_t|y_{t-1}] = e^{\left(\mu + \frac{\sigma^2}{2}\right)} ,\\ &\VAR[y_t|y_{t-1}] = \left(e^{\sigma ^{2}}-1\right)e^{\left(2\mu +\sigma ^{2}\right)} 
\end{align*}
Score calculation
\begin{align*}
\nabla_{t} &= \begin{bmatrix} 
\frac{\partial \ln p(y_t|y_{t-1}, \mu, \sigma^2)}{\partial \mu}\\
\frac{\partial \ln p(y_t|y_{t-1}, \mu, \sigma^2)}{\partial \sigma^2}\\
\end{bmatrix}\\
\ln p(y_t|y_{t-1}, \mu, \sigma^2) &= -\ln \left(y_t (1 - y_t)\right) - \frac{1}{2} \ln 2\pi\sigma^2 - \frac{(\text{logit}\left(y_t\right) - \mu)^2}{2\sigma^2} \\
\nabla_{t}^{\mu} = \frac{\partial \ln p(y_t|y_{t-1}, \mu, \sigma^2)}{\partial \mu} &= \frac{\text{logit}\left(y_t\right) - \mu}{\sigma^2}\\
\nabla_{t}^{\sigma^2} = \frac{\partial \ln p(y_t|y_{t-1}, \mu, \sigma^2)}{\partial \sigma^2} &= \frac{-1}{2\sigma^2}\left(1 - \frac{\text{logit}\left(y_t\right) - \mu)^2}{\sigma^2}\right)
\end{align*}

\subsection{Lognormal}
Density function
\begin{align*}
    &p(y_t|y_{t-1}, \mu, \sigma^2) = \frac{1}{y_t{\sqrt {2\pi\sigma^2}}}\ e^{ \left(-{\frac {\left(\ln y_t-\mu \right)^{2}}{2\sigma ^{2}}}\right)}, \quad \mu \in \mathbb{R}, \quad \sigma^2 \in \mathbb{R}^+\\
    &\E[y_t|y_{t-1}] = e^{\left(\mu + \frac{\sigma^2}{2}\right)} ,\\ &\VAR[y_t|y_{t-1}] = \left(e^{\sigma ^{2}}-1\right) e^{\left(2\mu +\sigma ^{2}\right)}  
\end{align*}
Score calculation
\begin{align*}
\nabla_{t} &= \begin{bmatrix} 
\frac{\partial \ln p(y_t|y_{t-1}, \mu, \sigma^2)}{\partial \mu}\\
\frac{\partial \ln p(y_t|y_{t-1}, \mu, \sigma^2)}{\partial \sigma^2}\\
\end{bmatrix}\\
\ln p(y_t|y_{t-1}, \mu, \sigma^2) &= -\ln y_t - \frac{1}{2} \ln 2\pi\sigma^2 - \frac{(\ln y_t - \mu)^2}{2\sigma^2} \\
\nabla_{t}^{\mu} = \frac{\partial \ln p(y_t|y_{t-1}, \mu, \sigma^2)}{\partial \mu} &= \frac{\ln y_t - \mu}{\sigma^2}\\
\nabla_{t}^{\sigma^2} = \frac{\partial \ln p(y_t|y_{t-1}, \mu, \sigma^2)}{\partial \sigma^2} &= \frac{-1}{2\sigma^2}\left(1 - \frac{(\ln y_t - \mu)^2}{\sigma^2}\right)
\end{align*}

\subsection{Negative binomial}
Density function
\begin{align*}
    &p(y_t|y_{t-1}, r, p) = \frac{\Gamma(y_t+r)}{y_t!\Gamma(r)}p^r\left(1-p\right)^{y_t}, \quad r \in \mathbb{R}^+, \quad p \in [0, 1]\\
    &\E[y_t|y_{t-1}] = \frac{pr}{1-p},\\ &\VAR[y_t|y_{t-1}] = \frac{pr}{(1-p)^2}
\end{align*}
Score calculation
\begin{align*}
\nabla_{t} &= \begin{bmatrix} 
\frac{\partial \ln p(y_t|y_{t-1}, r, p)}{\partial r}\\
\frac{\partial \ln p(y_t|y_{t-1}, r, p)}{\partial p}\\
\end{bmatrix}\\
\ln p(y_t|y_{t-1}, r, p) &= \ln \Gamma(y_t + r) - \left(\ln y_t! \Gamma(r)\right) + r\ln p + y_t \ln (1-p) \\
\nabla_{t}^{r} = \frac{\partial \ln p(y_t|y_{t-1}, r, p)}{\partial r} &= \psi(y_t + r) - \psi(r) + \ln p\\
\nabla_{t}^{p} = \frac{\partial \ln p(y_t|y_{t-1}, r, p)}{\partial p} &= \frac{r}{p} - \frac{k}{1-p}
\end{align*}

\subsection{Normal}
Density function
\begin{align*}
    &p(y_t|y_{t-1}, \mu, \sigma^2) = \frac{1}{\sqrt{2\pi\sigma^2}}e^{ \left({\frac{-(y_t - \mu)^2}{2\sigma^2}}\right)} , \quad \mu \in \mathbb{R}, \quad \sigma^2 \in \mathbb{R}^+\\
    &\E[y_t|y_{t-1}] = \mu,\\ &\VAR[y_t|y_{t-1}] = \sigma^2
\end{align*}
Score calculation
\begin{align*}
\nabla_{t} &= \begin{bmatrix} 
\frac{\partial \ln p(y_t|y_{t-1}, \mu, \sigma^2)}{\partial \mu}\\
\frac{\partial \ln p(y_t|y_{t-1}, \mu, \sigma^2)}{\partial \sigma^2}\\
\end{bmatrix}\\
\ln p(y_t|y_{t-1}, \mu, \sigma^2) &= \frac{-1}{2} \ln 2\pi\sigma^2 - \frac{(y_t - \mu)^2}{2\sigma^2} \\
\nabla_{t}^{\mu} = \frac{\partial \ln p(y_t|y_{t-1}, \mu, \sigma^2)}{\partial \mu} &= \frac{y_t - \mu}{\sigma^2}\\
\nabla_{t}^{\sigma^2} = \frac{\partial \ln p(y_t|y_{t-1}, \mu, \sigma^2)}{\partial \sigma^2} &= \frac{-1}{2\sigma^2}\left(1 - \frac{(y_t - \mu)^2}{\sigma^2}\right)
\end{align*}

\subsection{Poisson}
Density function
\begin{align*}
    &p(y_t|y_{t-1}, \lambda) = \frac{e^{-\lambda}\lambda^{y_t}}{y_t!}, \quad \lambda \in \mathbb{R}^+ \\ 
    &\E[y_t|y_{t-1}] = \lambda,\\ &\VAR[y_t|y_{t-1}] = \lambda
\end{align*}
Score calculation
\begin{align*}
\nabla_{t} &= \frac{\partial \ln p(y_t|y_{t-1}, \lambda)}{\partial \lambda}\\
\ln p(y_t|y_{t-1}, \lambda) &= -\lambda + y_t \ln \lambda - \ln y_t!\\
\nabla_{t}^{\lambda} = \frac{\partial \ln p(y_t|y_{t-1}, \lambda)}{\partial \lambda} &= \frac{y_t - \lambda}{\lambda}
\end{align*}

\subsection{Student's $t$}
Density function
\begin{align*}
    &p(y_t|y_{t-1}, \nu) = \frac{1}{\sqrt{\nu}B\left(\frac{1}{2},\frac{\nu}{2}\right)}\left(1 + \frac{y_t ^2}{\nu}\right)^{\frac{-\nu + 1}{2}}, \quad \nu \in \mathbb{R}^+\\
    &\E[y_t|y_{t-1}] = 0,\\ &\VAR[y_t|y_{t-1}] = \sigma^2 \frac{\nu}{\nu -2} \text{ for $\nu > 2$, $\infty$ for $1 < \nu \leq 2$, undefined otherwise.} 
\end{align*}
Score calculation
\begin{align*}
\nabla_{t} &= \begin{bmatrix} 
\frac{\partial \ln p(y_t|y_{t-1},\nu)}{\partial \nu}
\end{bmatrix}\\
\ln p(y_t|y_{t-1}, \nu) &= -\frac{1}{2}\ln \nu - \ln B\left(\frac{1}{2},\frac{\nu}{2}\right) - \left(\frac{\nu+1}{2}\right)\ln\left(1 + \frac{y_t^2}{\nu}\right) \\
\nabla_{t}^{\nu} = \frac{\partial \ln p(y_t|y_{t-1}\nu)}{\partial \nu} &= \frac{1}{2}\left(\frac{\left(\nu + 1\right)y_t^2}{\nu y_t^2 + \nu^2} - \frac{1}{\nu} -\ln\left(\frac{y_t ^2}{\nu} + 1\right) + \psi\left(\frac{\nu + 1}{2}\right)-\psi\left(\frac{\nu}{2}\right)\right)\\
\end{align*}

\subsection{Student's $t$ with Location and Scale}
Density function
\begin{align*}
    &p(y_t|y_{t-1}, \mu, \sigma^2, \nu) = \frac{1}{\sqrt{\sigma^2\nu}B\left(\frac{1}{2},\frac{\nu}{2}\right)}\left(1 + \frac{\left(y_t-\mu \right)^2}{\sigma^2\nu}\right)^{\frac{-\nu + 1}{2}}, \quad \mu \in \mathbb{R}, \quad \sigma^2 \in \mathbb{R}^+ \quad \nu \in \mathbb{R}^+\\
    &\E[y_t|y_{t-1}] = \mu,\\ &\VAR[y_t|y_{t-1}] = \sigma^2 \frac{\nu}{\nu -2} \text{ for $\nu > 2$, $\infty$ for $1 < \nu \leq 2$, undefined otherwise.} 
\end{align*}
Score calculation
\begin{align*}
\nabla_{t} &= \begin{bmatrix} 
\frac{\partial \ln p(y_t|y_{t-1}, \mu, \sigma^2, \nu)}{\partial \mu}\\
\frac{\partial \ln p(y_t|y_{t-1}, \mu, \sigma^2, \nu)}{\partial \sigma^2}\\
\frac{\partial \ln p(y_t|y_{t-1}, \mu, \sigma^2, \nu)}{\partial \nu}\\
\end{bmatrix}\\
\ln p(y_t|y_{t-1}, \mu, \sigma^2, \nu) &= -\frac{1}{2}\ln \nu\sigma^2 - \ln B\left(\frac{1}{2},\frac{\nu}{2}\right) - \left(\frac{\nu+1}{2}\right)\ln\left(1 + \frac{\left(y_t-\mu \right)^2}{\sigma^2\nu}\right) \\
\nabla_{t}^{\mu} = \frac{\partial \ln p(y_t|y_{t-1}, \mu, \sigma^2, \nu)}{\partial \mu} &= \frac{\left(\nu + 1\right)\left(y_t - \mu\right)}{\left(y_t - \mu\right)^2 + \sigma^2\nu}\\
\nabla_{t}^{\sigma^2} = \frac{\partial \ln p(y_t|y_{t-1}, \mu, \sigma^2, \nu)}{\partial \sigma^2} &= -\nu\frac{\sigma^2-\left(y_t - \mu\right)^2}{2\sigma^2\left(\nu\sigma^2+\left(y_t - \mu\right)^2\right)}\\
\nabla_{t}^{\nu} = \frac{\partial \ln p(y_t|y_{t-1}, \mu, \sigma^2, \nu)}{\partial \nu} &= \frac{1}{2}\left(\frac{\left(\nu + 1\right)\left(y_t - \mu\right)^2}{\nu\left(y_t - \mu\right)^2 + \sigma^2\nu^2} - \frac{1}{\nu} -\ln\left(\frac{\left(y_t - \mu\right)^2}{\sigma^2\nu} + 1\right)\right) +\\
&\quad \frac{1}{2}\left(\psi\left(\frac{\nu + 1}{2}\right)-\psi\left(\frac{\nu}{2}\right)\right)\\
\end{align*}

\subsection{Weibull}
Density function
\begin{align*}
    &p(y_t|y_{t-1},\lambda,k) =
\begin{cases}
\frac{k}{\lambda}\left(\frac{y_t}{\lambda}\right)^{k-1}e^{\left(-\frac{y_t}{\lambda}\right)^k} & x\geq0 ,\\
0 & x<0,
\end{cases} \quad \lambda \in \mathbb{R}^+, k \in \mathbb{R}^+\\
    &\E[y_t|y_{t-1}] = \lambda \Gamma\left(1 + 1/k\right),\\ &\VAR[y_t|y_{t-1}] = \lambda^2\left[\Gamma\left(1+\frac{2}{k}\right) - \left(\Gamma\left(1+\frac{1}{k}\right)\right)^2\right]
\end{align*}
Score calculation
\begin{align*}
\nabla_{t} &= \begin{bmatrix} 
\frac{\partial \ln p(y_t|y_{t-1}, \lambda, k)}{\partial k}\\
\frac{\partial \ln p(y_t|y_{t-1}, \lambda, k)}{\partial \lambda}\\
\end{bmatrix}\\
\ln p(y_t|y_{t-1}, \lambda, k) &= \ln k + (k - 1) \ln y_t - k\ln \lambda - \left(\frac{y_t}{\lambda}\right)^{k}\\
\nabla_{t}^{\lambda} = \frac{\partial \ln p(y_t|y_{t-1}, \lambda, k)}{\partial \lambda} &= \frac{k}{\lambda}\left(\left(\frac{y_t}{\lambda}\right)^k - 1\right)\\
\nabla_{t}^{k} = \frac{\partial \ln p(y_t|y_{t-1}, \lambda, k)}{\partial k} &= \frac{1}{k} + \ln\left(\frac{y_t}{\lambda}\right)\left(1 - \left(\frac{y_t}{\lambda}\right)^k\right)
\end{align*}

\newpage

\section{Computational details}

The results in this paper were obtained using
\proglang{Julia}~1.4.2 and \pkg{ScoreDrivenModels}~v0.1.2. Figure \ref{fig:ANE_scenarios} was generated with 
\pkg{Plots.jl}~v1.4.3. In order to guarantee that the results of the examples are exactly the same as the ones reported, we recommend the interested user to \code{activate} the \code{Project.toml} in the repository's example folder using Julia's \pkg{Pkg} manager.

\end{appendix}


\end{document}